\documentclass[twocolumn,amsmath,amssymb,prl,aps,superscriptaddress,10pt]{revtex4-1}
\usepackage[utf8]{inputenc}
\usepackage{graphicx}% Include figure files
\usepackage{dcolumn}% Align table columns on decimal point
\usepackage{bm}% bold math
\usepackage{epsfig}
\usepackage{booktabs}
\usepackage[dvipsnames]{xcolor}
\usepackage[normalem]{ulem}
\usepackage{subfigure}
\usepackage{enumitem}
\usepackage{epstopdf}
\usepackage{hyperref}
\usepackage{appendix}
\usepackage{ulem}
\hypersetup{colorlinks=true,linkcolor=blue,citecolor=blue,menucolor=black,urlcolor=blue,filecolor=blue}

\newcommand{\der}{\mathrm{d}}
\newcommand{\be}{\begin{equation}}
\newcommand{\ee}{\end{equation}}
\newcommand{\bea}{\begin{eqnarray}}
\newcommand{\eea}{\end{eqnarray}}
\newcommand{\bean}{\begin{eqnarray*}}
\newcommand{\eean}{\end{eqnarray*}}

\renewcommand{\sout}{\bgroup \color[rgb]{1,0,0} \ULdepth=-.5ex \ULset}

\newcommand{\itp}{\affiliation{CAS Key Laboratory of Theoretical Physics, Institute of Theoretical Physics,\\ Chinese Academy of Sciences, Beijing 100190, China}}
\newcommand{\ucas}{\affiliation{School of Physical Sciences, University of Chinese Academy of Sciences, Beijing 100049, China}}

\begin{document}

\title{Graphic Method for Arbitrary $n$-body Phase Space}

\author{Hao-Jie Jing} \email{jinghaojie@itp.ac.cn}
\itp
\ucas

\author{Chao-Wei Shen} \email{shencw@ucas.ac.cn}
\ucas

\author{Feng-Kun Guo} \email{fkguo@itp.ac.cn}
\itp
\ucas

\date{\today}

\maketitle

In high energy physics, decay widths and cross sections are among the most important physical observables, and the computation of each of them contains a phase space integration. Therefore, the phase space integration is essential to connect theoretical calculations with experimental observations. It also enters into the calculation of the imaginary part of a physical amplitude through unitarity.
Usually, in experiments, it is important for understanding the structure of particle spectrum whether there is a nontrivial structure in the invariant mass distribution. For instance, resonances, such as excited hadrons and the Higgs particle, are often found as peaks in the invariant mass distributions of certain final state particles.
Thus, a key problem of the phase space integration is to get the formula of the phase space element expressed in terms of any given invariant masses for an $n$-body system.
When there are only two or three particles in the final state, the corresponding phase space integration is relatively easy, as given in, e.g., the chapter of Kinematics in the Review of Particle Physics~\cite{Zyla:2020zbs}.
When the number of final state particles is larger than 3, the phase space integration becomes much more involved.
In this paper, we propose a novel method based on graphics, which can not only give the phase space formula of any given invariant masses intuitively in general $D$-dimensional spacetime, but also greatly simplifies the calculation just as what Feynman diagrams do in calculating scattering amplitudes.

The graphic method proposed here is applicable in the case of general $D$-dimensional spacetime, which is useful when considering dimensional regularization (e.g., in the soft-gluon resummation in Ref.~\cite{Forte:2002ni}) and quantum field theory in arbitrary dimensions. Because the four-dimensional spacetime is more commonly used in calculations of high-energy physics, we only introduce the four-dimensional case here (for the general case of $D$-dimensional spacetime, see supplemental materials). The $n$-body phase space element is~\cite{Zyla:2020zbs}
\begin{equation}
\mathrm{d}\Phi_n(m;m_1,\ldots,m_n) = \delta^4(p-\sum_{i=1}^n p_i) \prod_{j=1}^n \frac{\mathrm{d}^3\mathbf{p}_j}{(2\pi)^3 2p^0_j},
\label{eq_NBodyPSE}
\end{equation}
where $p$ and $p_i$ are the four-momenta of the initial state and the $i$-th particle in the final state, respectively, and they satisfy the on-shell conditions $p^2=m^2$ and $p_i^2=m_i^2\,(i=1,\cdots,n)$ with $m$ and $m_i$ the corresponding particle masses. The final state particle masses $m_i$ can be either vanishing or finite.
One notices that the phase space element can be written in an explicitly Lorentz-invariant manner as
\begin{equation}
    \frac{\mathrm{d}^3\mathbf{p}_j}{(2\pi)^3 2p^0_j} = \frac{\mathrm{d}^4 p_j}{(2\pi)^3} \delta(p_j^2 - m_j^2)\theta(p_j^0).
    \label{eq_pselement}
\end{equation}

After integrating out the order-4 Dirac $\delta$-function in Eq.~\eqref{eq_NBodyPSE}, which represents the energy-momentum conservation, $3n-4$ integral variables are left.
In order to get invariant mass distributions, variable transformations from three momenta to invariant masses are needed. The final state in the momentum space is just like a rigid body when the magnitude of the three-momentum for each particle and the relative angle between each two momenta are fixed. One can make a global Euler rotation of the final state, which does not change the value of any invariant mass.
Thus, when considering only the scalar products of the final state momenta, among all the $3n-4$ variables, at most $3n-7$ ($n\geq3$) variables can be replaced by invariant masses, and the other three variables can be replaced by Euler angles $\alpha,\beta$ and $\gamma$.

\begin{figure}[tb]
  \centering
%   \subfigure[]{
%     \includegraphics[width=0.65\linewidth]{recursive}\label{fig_graph_a}
%     }\\
%   \subfigure[]{
%     \includegraphics[width=0.47\linewidth]{1tonDec_chaintree}\label{fig_graph_b}
%     }\hfill
%   \subfigure[]{
%     \includegraphics[width=0.47\linewidth]{1to3Dec_loop}\label{fig_graph_c}
    % }
    \includegraphics[width=\linewidth]{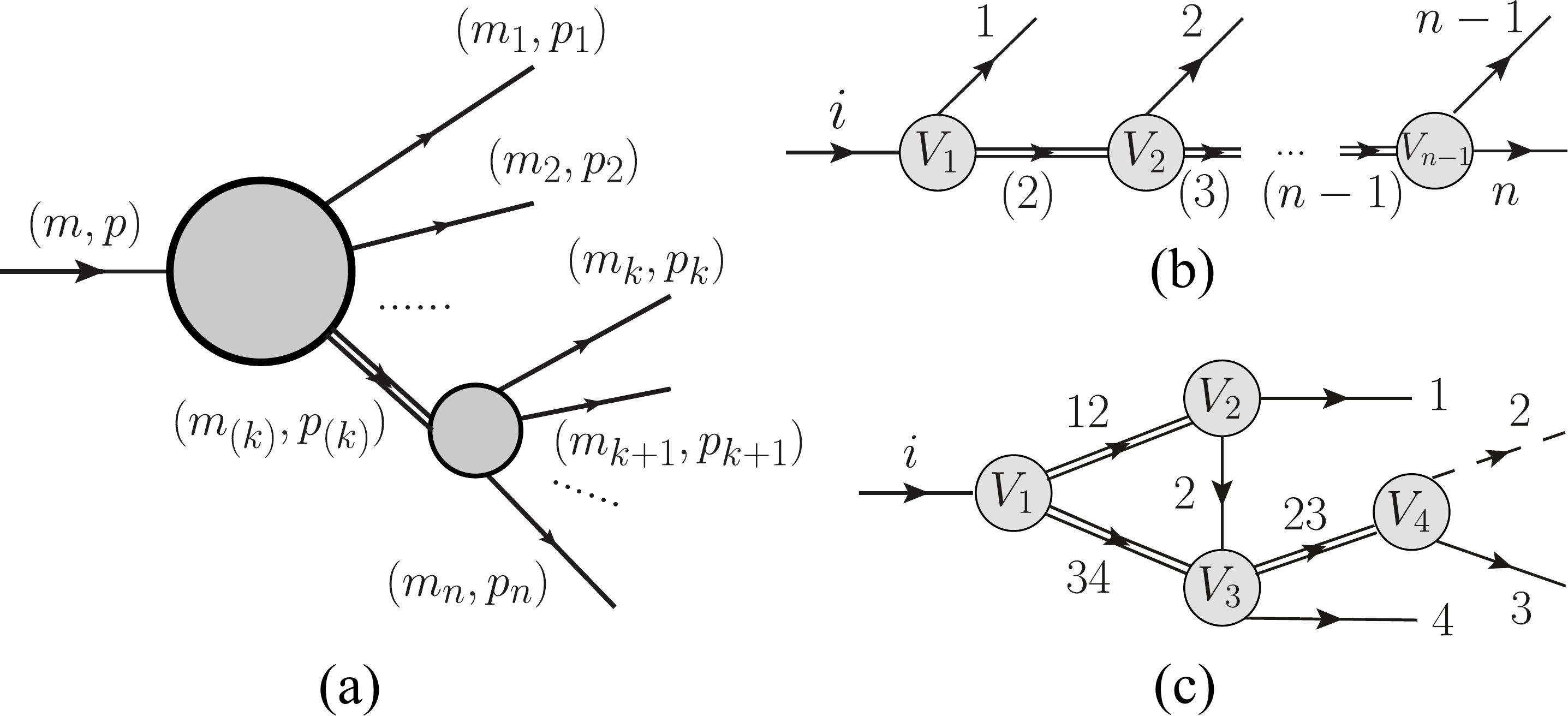} %\label{fig_graph_c}
    \caption{Phase space elements expressed by the graphic method: (a) graphic representation of Eq.~\eqref{eq_recursive}; (b) chain tree diagram of $n$-body phase space; (c) 1-loop diagram of 4-body phase space. Notice that these graphs should not be understood as Feynman diagrams. }
    \label{fig_graph}
\end{figure}

The graphic method is based on the recursive relation
\begin{align}
    \der\Phi_n(m;m_1,\ldots,m_n) =&\, \der\Phi_k(m;m_1,\ldots,m_{(k)} ) (2\pi)^3 \der m_{(k)}^2 \nonumber\\
    & \times\der\Phi_{n-k+1}(m_{(k)};m_k,\ldots,m_n),
    \label{eq_recursive}
\end{align}
where $m_{(k)}^2 = (p_k +\ldots+p_n)^2$ with $k<n$. This relation can be represented graphically as Fig.~\ref{fig_graph}(a).
We find that the generic $n$-body phase space expression in terms of invariant masses can be obtained using the following drawing rules:
\begin{enumerate}[leftmargin=*,label={(\arabic*)}]
\item A single particle is represented by a single line:\\
      \begin{tabular}{ll}
      \parbox[c]{3.8cm}{\includegraphics[height=7mm]{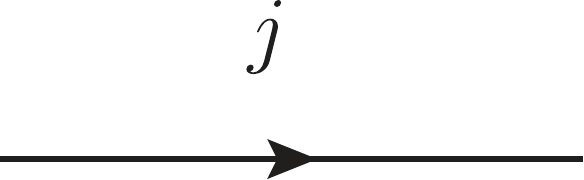}} & 1;
      \end{tabular}

      a multi-particle system is represented by a double line:\\
      \begin{tabular}{ll}
      \parbox[c]{3.8cm}{\includegraphics[height=7mm]{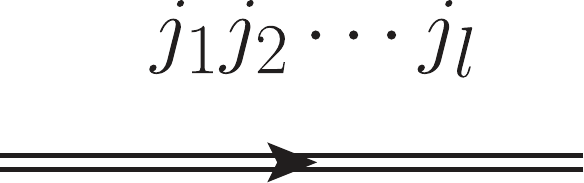}} & $(2\pi)^3\der m_{j_1j_2\cdots j_l}^2$;
      \end{tabular}

      an $l$-body phase space element is represented by a vertex:\\[2mm]
      \begin{tabular}{ll}
      \parbox[c]{3.8cm}{\includegraphics[height=1.8cm]{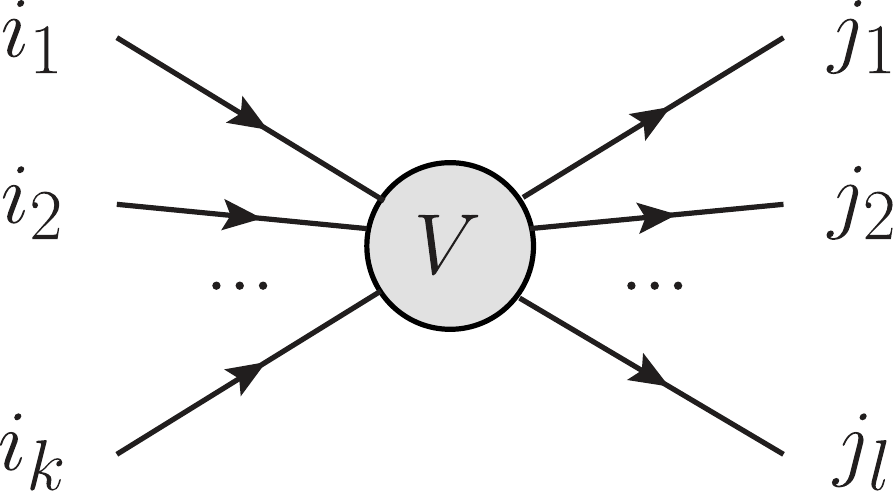}} & $\der\Phi_l(m_{i_1\cdots i_k}; m_{j_1},\ldots,m_{j_l})$,
      \end{tabular}\\[2mm]
      where the lines can be either single or double lines and $m_{i_1\cdots i_k}^2 = (p_{j_1}+\ldots+p_{j_l})^2$.
\label{Rule_1}

\item A single line can be internal or external, and a double line can only be internal.
\label{DoubleIn}

\item There is one and only one route of double lines between any two vertices.
\label{BridgeRule}

\item If there are duplicate single lines for the same particle in the whole diagram, only one can be kept, and the rest are represented by dashed single lines:\\
      \begin{tabular}{ll}
      \parbox[c]{3.8cm}{\includegraphics[height=7mm]{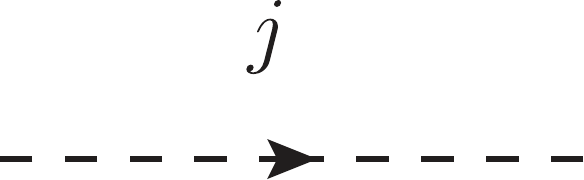}} & $\dfrac{(2\pi)^3}{\mathrm{d}^4p_j^{}\delta(p_j^2-m_j^2)\theta(p_j^0)}$.
      \end{tabular}
\label{DashedLine}
This, following from Eq.~\eqref{eq_pselement}, is to cancel out one copy of the phase space element of the particle which is double-counted in the integral measures (vertices) involving it.

\item Invariant masses for all double lines in the whole diagram must be independent (for instance, $m_{12}$, $m_{23}$ and $m_{13}$ for a 3-body system are not independent as they satisfy $m_{12}^2+m_{23}^2+m_{13}^2 = m^2 + \sum_{i=1}^3m_i^2$).
\end{enumerate}

From the above drawing rules, one can find the following topological rules:
\begin{itemize}%[leftmargin=*]
%\begin{enumerate}[leftmargin=*,label={(\arabic*)}]
    \item[(i)] using Rule~\ref{DoubleIn} and Rule~\ref{BridgeRule}, one finds that the number of vertices $v$ and that of double lines $d$ are related as $v=d+1$, which ensures that a Dirac $\delta$-function for the overall energy-momentum conservation of the whole diagram exists;
    \item[(ii)] the number of final state particles $n$, the number of double lines $d$, the number of internal single (solid and dashed) lines $l$ and the number of outgoing lines for each vertex $v_j$ are related as $n+d+l=\sum_{j=1}^{d+1}v_j$;
    \item[(iii)] using Rule~\ref{DashedLine}, the number of dashed lines equals the number of internal single lines $l$;
    \item[(iv)] using Euler's formula: $v-(l+d)+L$=1, where $L$ is the number of loops in a diagram, one finds $L=l$.
\end{itemize}
%\end{enumerate}

A diagram with $v_j=2$ for all vertices is called a {\em complete-expansion diagram}, such as Fig.~\ref{fig_graph}(b) and \ref{fig_graph}(c); otherwise, it is called an {\em incomplete-expansion diagram}, such as Fig.~\ref{fig_graph}(a).
Because an incomplete-expansion diagram can be further expanded into a complete-expansion diagram, we shall only discuss the complete-expansion diagrams in the following. For a complete-expansion diagram, one has $d-l=n-2$.
A diagram without an internal single line, i.e. $l=0$, is called a {\em tree diagram}, and a diagram with $l>0$ is called an {\em $l$-loop diagram}. For a tree diagram, one finds $d=n-2$, which means that we can get an $n$-body phase space element with $n-2$ invariant masses as the integral variables. From the above discussions, at most $3n-7$ variables can be replaced by invariant masses in the $n$-body ($n\geq3$) phase space element.
Then, if $d$ takes its maximal value $3n-7$, one will get $l=2n-5$, which means that a $(2n-5)$-loop diagram corresponds to an $n$-body phase space element with $3n-7$ invariant masses as the integral variables.

Using the graphic method, the $n$-body phase space element can be easily decomposed into the product of many 2-body phase space elements.
All possible forms of the 2-body phase space element are listed in supplemental materials, and here we take the solid angle as the integral variables. The 2-body phase space element in any reference frame can be written as
\begin{align}
    \der\Phi_2(m;m_1,m_2) = &\sum_{|\mathbf{p}_1|}\frac{\der\Omega_1}{(2\pi)^6} \frac{|\mathbf{p}_1|^2}{4|(p^0|\mathbf{p}_1|-p_1^0|\mathbf{p}|\cos{\theta_{01}})|} \nonumber\\ &\times\theta\left(p^0-\sqrt{|\mathbf{p}_1|^2+m_1^2}\right),
    \label{eq_2body}
\end{align}
where $p,p_1$ and $\theta_{01}$ are the momentum of the initial state, the momentum of particle 1 in the final state and the relative angle between $\mathbf{p}$ and $\mathbf{p}_1$, respectively. The summation runs over all $|\mathbf{p}_1|$ solutions of the physical on-shell equations.
The effect of the Heaviside $\theta$-function is to limit the integration range to the physical region. One may omit it for simplicity, but needs to be careful with the integration range.
In the c.m. frame of the initial state,  $p=(p^0,\mathbf{p})=(m,\mathbf{0})$, and one gets
\begin{equation}
    \der\Phi_2(m;m_1,m_2) = \der\Omega_1\frac{|\mathbf{p}_1|}{(2\pi)^6 4m},
    \label{eq_1to2PSFcm}
\end{equation}
where $|\mathbf{p}_1|$ is the magnitude of the three-momentum of particle 1 in the c.m. frame of the initial state, and the integration region is given by $\cos{\theta_1}\in[-1,1]$ and $\phi_1\in[0,2\pi)$.

Using the graphic method and Eq.~\eqref{eq_2body}, an arbitrary $n$-body phase space can be easily expressed in integrations over any allowed invariant masses with the involved momenta being in any reference frame. One just needs to draw the graphs with these invariant masses as double lines following the rules presented above. Next, we give two examples depicted in Fig.~\ref{fig_graph} to show how to get the $n$-body $(n\geq3)$ phase space element by using this method. In these examples, we use the c.m. two-body phase space in Eq.~\eqref{eq_1to2PSFcm} for simplicity.

For Fig.~\ref{fig_graph}(b), which is called the chain tree diagram of $n$-body phase space, using the corresponding rules, one has the following building blocks:
\begin{equation}
\begin{split}
    &V_1:~ \der\Phi_2(m;m_1,m_{(2)}),\hspace{.3cm}V_2:~ \der\Phi_2(m_{(2)};m_2,m_{(3)}),~\ldots,\\ &V_{n-1}:~ \der\Phi_2(m_{(n-1)};m_{n-1},m_n),\\
    &``(2)\text{''}:~(2\pi)^3\der m^2_{(2)},\hspace{8mm}``(3)\text{''}:~(2\pi)^3\der m^2_{(3)},~\ldots,\\
    &``(n-1)\text{''}:~(2\pi)^3\der m^2_{(n-1)},
    \notag
\end{split}
\end{equation}
where $V_i$ and ``$(j)$'' are as labelled in the diagram.
Multiplying them together and using Eq.~\eqref{eq_1to2PSFcm}, an expression for the $n$-body phase space element can be easily obtained,
\begin{equation}
    \der\Phi_n(m;m_1,\ldots,m_n)
    =\frac{|{\mathbf p}_1|\der\Omega_1}{2^{n}(2\pi)^{3 n} m} \prod_{i=2}^{n-1} |{\mathbf p}_{i}|\der\Omega_i\der m_{(i)}^{},
\end{equation}
where $(|\mathbf{p}_i|, \Omega_i)$ is the three-momentum of the final-state particle $i$ in the c.m. frame of the ($i,i+1,\ldots,n$) particle system. The integration region for the invariant mass $m_{(i)}$ is $\left[\sum^{n}_{k=i}m_k,m_{(i-1)}-m_{i-1}\right](i=2,3,\ldots,n-1)$ with $m_{(1)}=m$. A much more lengthy derivation can be found in the appendix of Ref.~\cite{Jing:2019cbw} (see also Ref.~\cite{Forte:2002ni}).

Next, let us consider a four-body phase space which can be troublesome to derive in the conventional approach. In addition to the chain tree diagram, which reduces to the integration over two invariant masses and angular variables, the phase space sometimes needs to be expressed in terms of three invariant masses, which can be easily obtained from Fig.~\ref{fig_graph}(c) by applying the rules introduced above. The building blocks are
\be
\begin{split}
    &V_1:~ \der\Phi_2(m;m_{12},m_{34}),\hspace{.8cm} V_2:~ \der\Phi_2(m_{12};m_1,m_2),\\
    &V_{3}:~ \der\Phi_2(m_{234};m_4,m_{23}),\hspace{.5cm} V_{4}:~ \der\Phi_2(m_{23};m_2,m_3),\\
    &``12\text{''}:~(2\pi)^3\der m^2_{12},\quad ``34\text{''}:~(2\pi)^3\der m^2_{34},\\
    &``23\text{''}:~(2\pi)^3\der m^2_{23},\\
    &\text{dashed single line ``2''}:~ (2\pi)^{3}\left[\mathrm{d}^4p_2\delta(p_2^2-m_2^2)\theta(p^0_2)\right]^{-1}.
    \notag
\end{split}
\ee
Multiplying them together, using Eqs.~\eqref{eq_pselement} and \eqref{eq_1to2PSFcm}, and integrating out the remaining $\delta$ function, one gets
\begin{align}
    &\der\Phi_4(m;m_1,m_2,m_3,m_4) \notag\\
    =\,& \frac{|\mathbf{p}'_{12}||\mathbf{p}''_1|}{(2\pi)^{12} 2^7m m_{12}m_{234}|\mathbf{p}^*_2|}\der m_{12}^2\der m_{34}^2\der m_{23}^2\der\Omega_{12}'\der\Omega_1''\der\phi_4^*  \notag\\
    &\times\theta_{[-1,1]}\left(\frac{m_{23}^2+m_2^2-m_3^2-2p_{23}^{*0}p_2^{*0}}{2|\mathbf{p}_2^*||\mathbf{p}_4^*|}\right),
\end{align}
where $(|\mathbf{p}'_{12}|, \Omega_{12}')$ is the three-momentum of the final-state (1,2) particle system in the c.m. frame of the initial state, $(|\mathbf{p}''_{1}|, \Omega_{1}'')$ is the three-momentum of particle 1 in the c.m. frame of particles 1 and 2, and the quantities labelled by a ``$^*$'' are defined in the c.m. frame of the (2,3,4) particle system in the final state. The invariant mass of particles 2, 3 and 4, $m_{234}$, is a function of $m_{12}$, $m_{34}$, $\Omega_{12}'$ and $\Omega_1''$.
The integration regions of $m_{12}$ and $m_{34}$ are $[m_1+m_2,m-m_3-m_4]$ and $[m_3+m_4,m-m_{12}]$, respectively, and the integration region of $m_{23}$ is limited by the boxcar function, defined as $\theta_{[-1,1]}(x)=1$ for $x\in[-1,1]$ and 0 otherwise.

More details and additional examples can be found in the  supplemental materials.

{\it Summary.}---In this article, we propose a graphic method which can greatly simplify the phase space calculation. The method is generic for evaluating an arbitrary $n$-body phase space in arbitrary spacetime dimensions. By combining the result of the 2-body phase space element with the graphic method, one can obtain the $n$-body phase space element in terms of any given invariant masses intuitively and efficiently: one simply follows the rules to draw the diagram containing double lines for these invariant masses. The involved momenta can be in any reference frame by using the two-body phase space element in that frame given here. In high-energy physics, the phase space element is an essential part in computing reaction cross sections and particle decay widths; in quantum field theory, it also enters into the calculation of the imaginary part of a physical amplitude through unitarity. A broad use of this method is foreseen.

\bigskip

This work is supported in part by the National Natural Science Foundation of China (NSFC) under Grants No.~11835015, No.~11947302, No.~11961141012 and No.~11621131001 (the Sino-German Collaborative Research Center CRC110 ``Symmetries and the Emergence of Structure in QCD"), by the Chinese Academy of Sciences (CAS) under Grants No.~XDB34030303 and No.~QYZDB-SSW-SYS013, and by the CAS Center for Excellence in Particle Physics (CCEPP).

\begin{widetext}
\begin{center}
\appendix{{\bf Supplemental Materials}}
\end{center}
\section{Material 1: $2$-body phase space element in $4$-dimensional spacetime}
All possible forms of the 2-body phase space element are listed in Table~\ref{tab_2bodyPS_results},
where the presence of an $\aleph_1$ means that the physical on-shell equations
\be
\left\{
\begin{aligned}
\hspace{.1cm}&p_1^2=(p_1^0)^2-|\mathbf{p}_1|^2=m_1^2, \\
\hspace{.1cm}&p\cdot p_1=p^0p_1^0-|\mathbf{p}||\mathbf{p}_1|\cos{\theta_{01}}=\frac{m^2+m_1^2-m_2^2}{2}
\end{aligned}
\right.
\label{eq_onshell}
\ee
admit an infinity of solutions in that special case, where $p,p_1$ and $\theta_{01}$ are the momentum of the initial state, the momentum of particle 1 in the final state and the relative angle between $\mathbf{p}$ and $\mathbf{p}_1$, respectively.

\begin{table}[h]
    \centering
    \caption{Integrand of the 2-body phase space element in 5 cases with different integral variables. A factor $(2\pi)^{-6}$ has been omitted in each integrand. The definitions of $p,p_1$ and $\theta_{01}$ can be found below Eq.~\eqref{eq_onshell}. The last three columns give the possible numbers of solutions of the physical on-shell equations of particles 1 and 2 in the final state when the corresponding two integral variables are fixed in any reference frame, where $\beta$ and $\beta_1^*$ are the velocity of the initial state and that of particle 1 in the rest frame of the initial state, respectively. $\aleph_1$ is the second transfinite number.}
    \begin{ruledtabular}
    \begin{tabular}{lcccc}
        Integral measure & Integrand & $\beta=0$ & $0<\beta<\beta_1^*$ & $\beta\geq\beta_1^*$\\
    \midrule[.5pt]
        $\der\cos{\theta_1}\der\phi_1$ & $\frac{|\mathbf{p}_1|^2}{4|(p^0|\mathbf{p}_1|-p_1^0|\mathbf{p}|\cos{\theta_{01}})|}$ & 1 & 1 & 1,2 \\
        $\der|\mathbf{p}_1|\der\phi_1$ & $\frac{|\mathbf{p}_1|}{4p_1^0|\mathbf{p}||\partial\cos{\theta_{01}}/\partial\cos{\theta_1}|}$ & $\aleph_1$ & 1,2 & 1,2 \\
        $\der p_1^0\der\phi_1$ & $\frac{1}{4|\mathbf{p}||\partial\cos{\theta_{01}}/\partial\cos{\theta_1}|}$ & $\aleph_1$ & 1,2 & 1,2 \\
        $\der|\mathbf{p}_1|\der\cos{\theta_1}$ & $\frac{|\mathbf{p}_1|}{4p_1^0|\mathbf{p}||\partial\cos{\theta_{01}}/\partial\phi_1|}$ & 1,$\aleph_1$ & 1,2,$\aleph_1$ & 1,2,$\aleph_1$ \\
        $\der p_1^0\der\cos{\theta_1}$ & $\frac{1}{4|\mathbf{p}||\partial\cos{\theta_{01}}/\partial\phi_1|}$ & 1,$\aleph_1$ & 1,2,$\aleph_1$ & 1,2,$\aleph_1$ \\
    \end{tabular}
    \end{ruledtabular}
    \label{tab_2bodyPS_results}
\end{table}

Here, we give an example to explain more about the presence of $\aleph_1$, which means that Eq.~\eqref{eq_onshell} admit an infinity of solutions. Let us consider the case with $(|\mathbf{p}_1|,\phi_1)$ being the integral variables. In the spherical coordinate system of the $\mathbf{p}_1$ space, the isoline of fixed $|\mathbf{p}_1|$ and $\phi_1$ is a semicircle ($\theta_1\in [0,\pi]$). Meanwhile, in the c.m. frame of the initial state ($\beta=0$), the solutions $\mathbf{p}_1$ of the physical on-shell equations correspond to a sphere centered at the origin. The isoline is completely on the sphere, which means that the on-shell equations are satisfied by every point on the isoline, and thus there are infinite ($\aleph_1$) solutions. In such a case, the corresponding 2-body phase space element cannot be used.

\begin{figure}[b]
    \centering
    \includegraphics[width=0.355\textwidth]{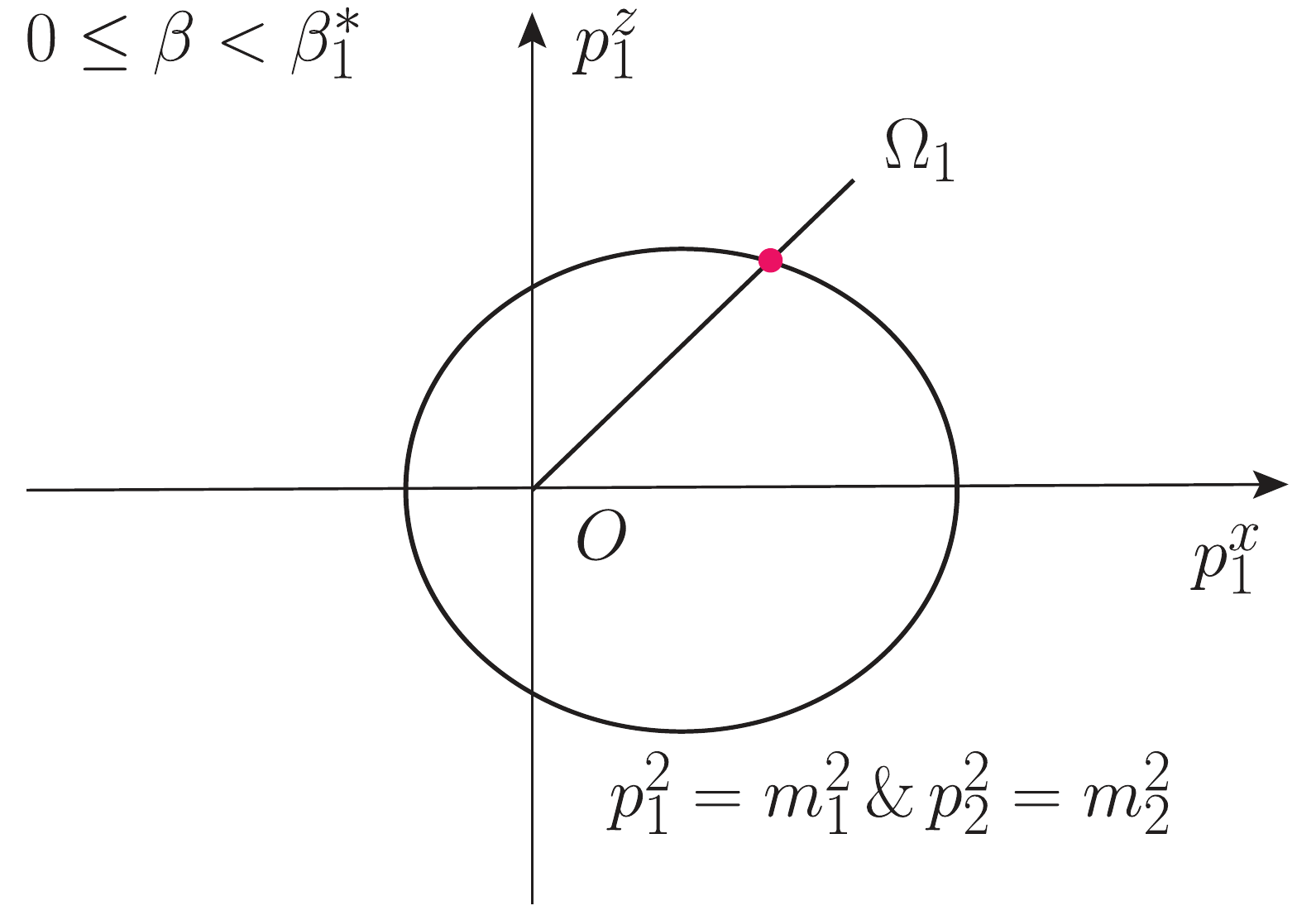}\hspace{.5cm}
    \includegraphics[width=0.365\textwidth]{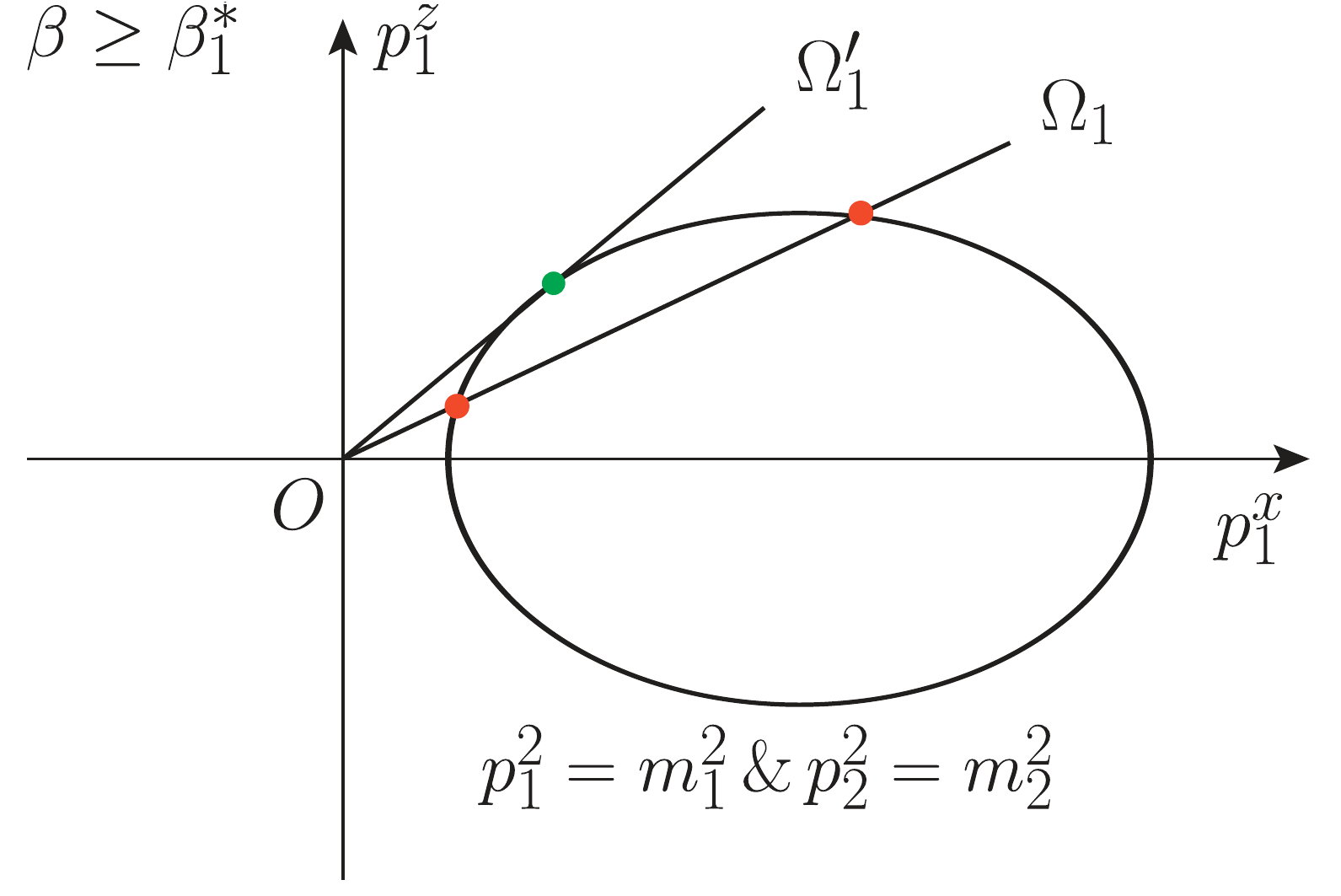}
    \caption{2-body phase space in the $xz$-plane of the momentum space of particle-1. Left: when $0\leq\beta<\beta_1^*$, the on-shell equations have one solution for each solid angle $\Omega_1$. Right: when $\beta\geq\beta_1^*$, the on-shell equations have one solution or two solutions with different magnitude of momentum $|\mathbf{p}_1|$ for each solid angle $\Omega_1$.}
    \label{fig:solution}
\end{figure}

In addition, let us give an example to explain why sometimes there are two or three numbers in Table~\ref{tab_2bodyPS_results}; for instance, when choosing $(\cos\theta_1,\phi_1)$ as the integral variables and $\beta\geq\beta^*_1$, Eq.~(1) admits either 1 or 2 solutions. Consider the solution of Eq.~(1) in final state momentum space, it is a sphere with a radius of $|\mathbf{p}_1^*|$, where $|\mathbf{p}_1^*|$ is the magnitude of momentum of particle-1 in the c.m. frame of the initial state, and the corresponding velocity is $\beta_1^*=|\mathbf{p}_1^*|/\sqrt{|\mathbf{p}_1^*|^2+m_1^2}$. If we make a boost of the initial state along the $x$-axis with a boost parameter $\beta$, the sphere will become a rotational ellipsoid with the first eccentricity being $\beta$, see Fig.~\ref{fig:solution}. When the velocity of the initial state $\beta$ is smaller than the velocity of particle-1 in the c.m. frame of the initial state, i.e., $0\leq\beta<\beta_1^*$, the origin is inside the ellipsoid, and one will get one solution for each solid angle $\Omega_1$. When the velocity of the initial state $\beta$ is equal to or larger than the velocity of particle-1 in the c.m. frame of decaying particle, i.e., $\beta\geq\beta_1^*$, the origin is on the ellipsoid or outside the ellipsoid; in this case, for each solid angle $\Omega_1$ one will get one solution or two solutions with different magnitudes of momentum $|\mathbf{p}_1|$.

\section{Material 2: Some examples of computing phase space by graphic method}

\begin{figure}[h]
  \centering
  \subfigure[]{
    \includegraphics[width=0.35\linewidth]{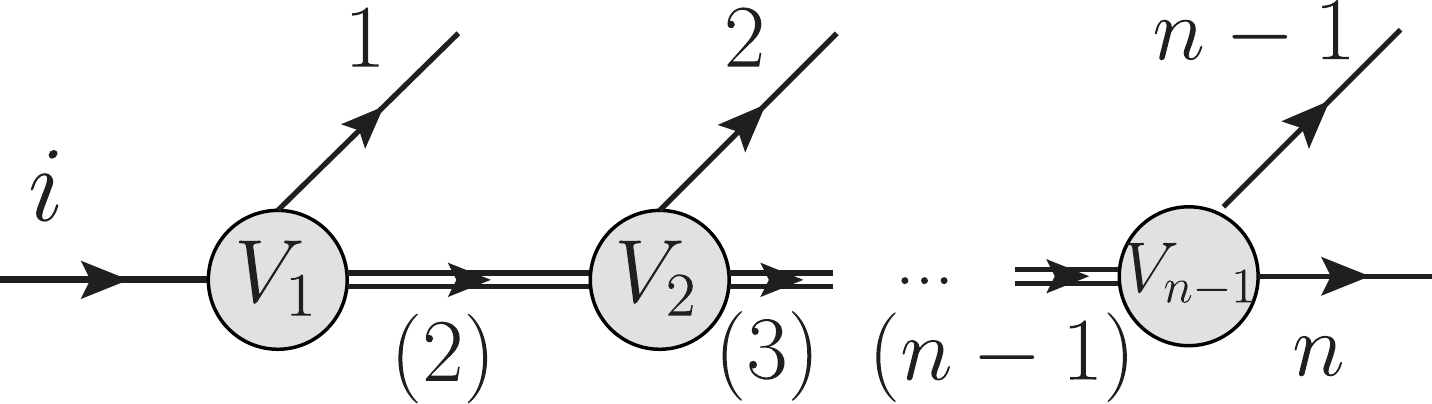}\label{fig_TwoDecEX_a}
    }\hspace{.5cm}
  \subfigure[]{
    \includegraphics[width=0.35\linewidth]{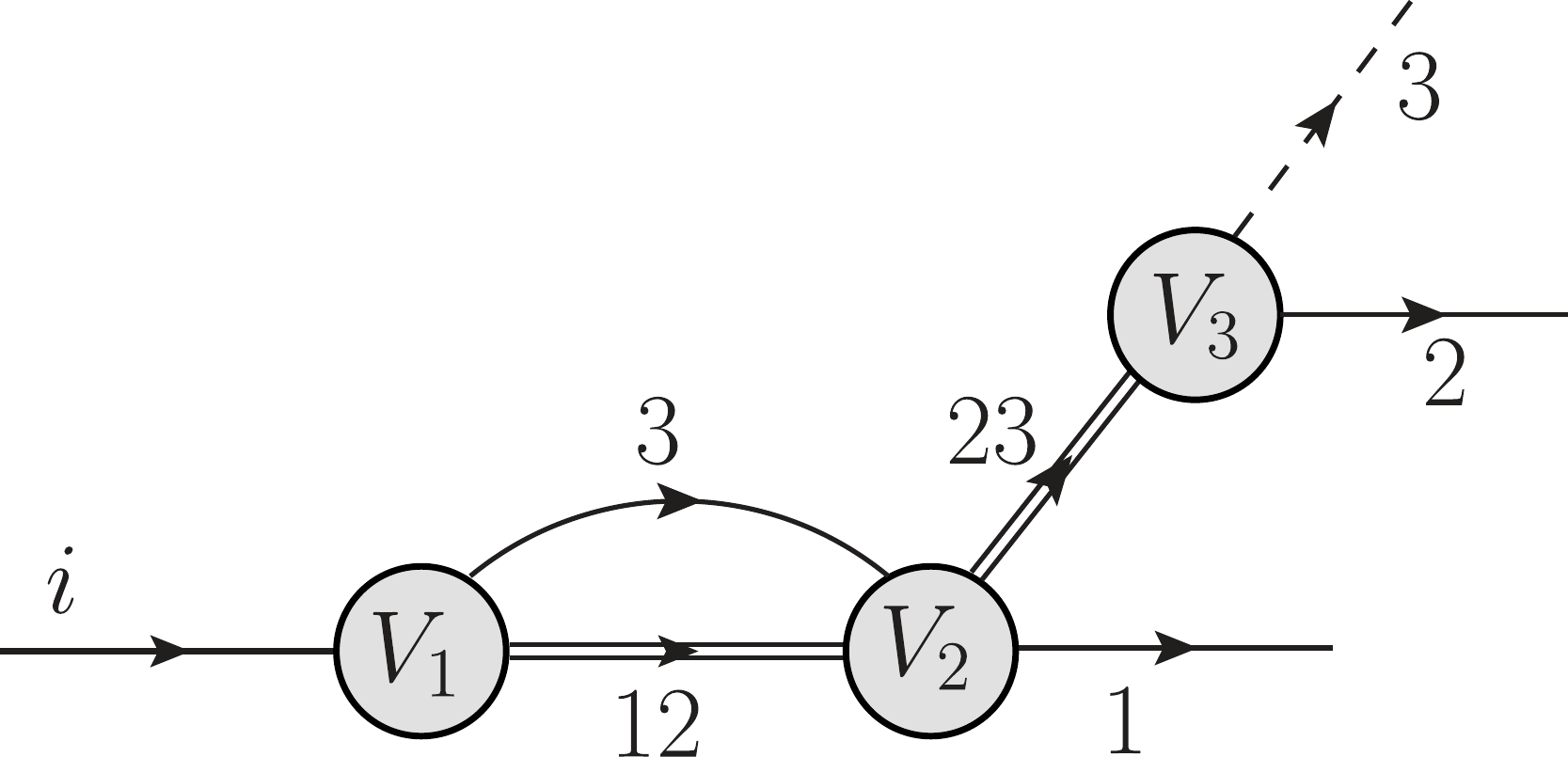}\label{fig_TwoDecEX_b}
    } \\
  \subfigure[]{
    \includegraphics[width=0.35\linewidth]{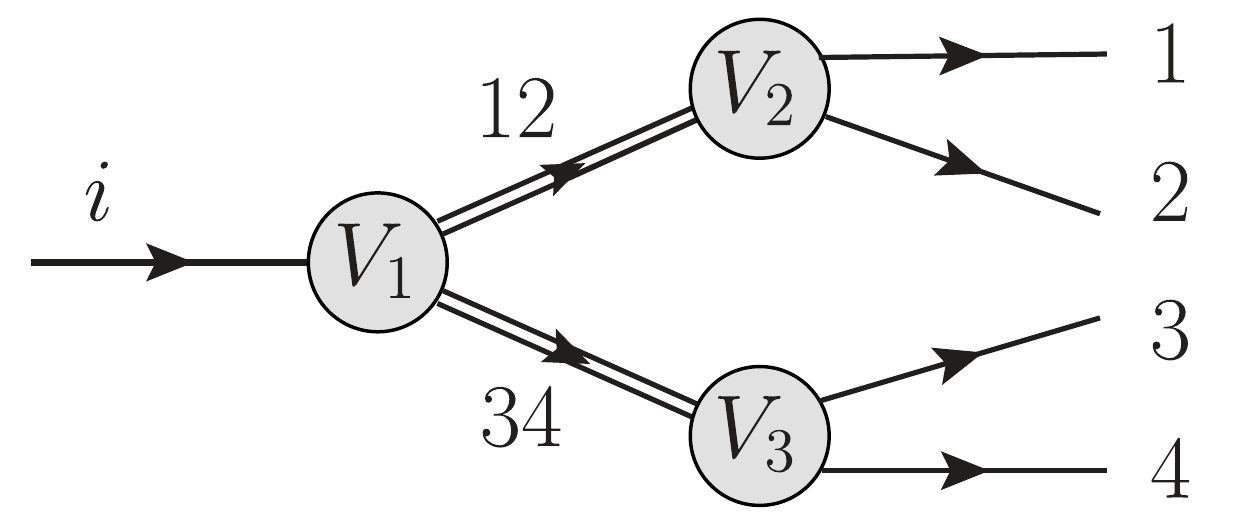}\label{fig_TwoDecEX_c}
    }\hspace{.5cm}
  \subfigure[]{
    \includegraphics[width=0.35\linewidth]{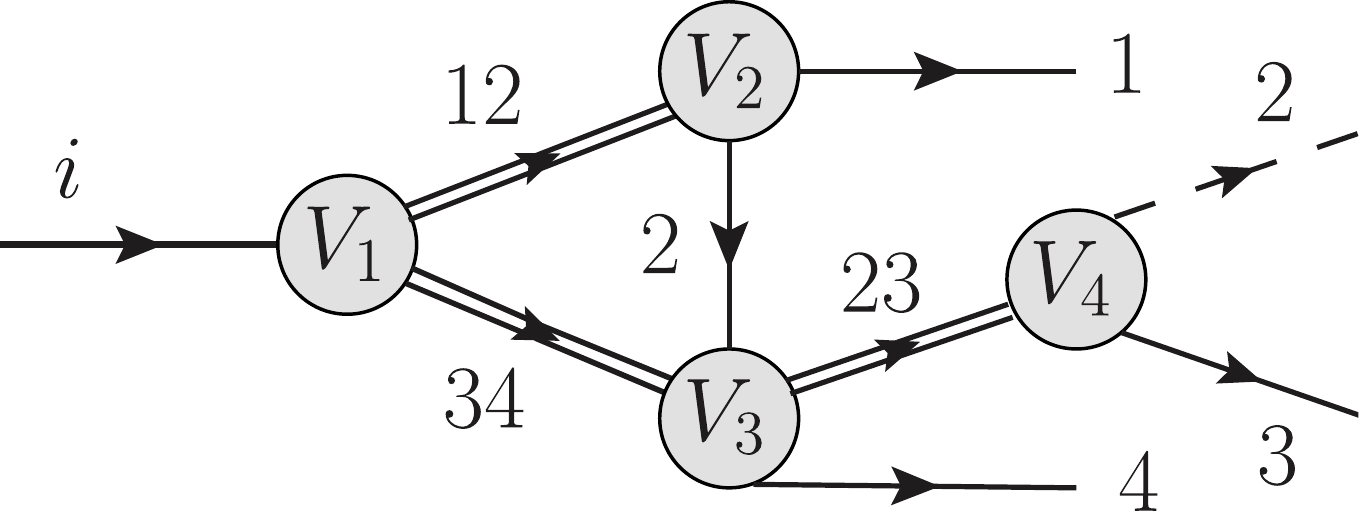}\label{fig_TwoDecEX_d}
    }
    \caption{Some examples of computing the phase space element by the graphic method: (a) is an $n$-body phase space; (b) is a 3-body phase space; (c) and (d) correspond to the 4-body phase space in terms of two different combinations of invariant masses. Notice that these graphs should not be understood as Feynman diagrams. }
    \label{fig_TwoDecEX}
\end{figure}

In this section, we present more details for the examples discussed in the main text and give additional examples.
In these examples, we use the following two-body phase space in the c.m. frame for simplicity.
\begin{equation}
    \der\Phi_2(m;m_1,m_2) = \der\Omega_1\frac{|\mathbf{p}_1|}{(2\pi)^6 4m},
    \label{eq_1to2PSFcm}
\end{equation}
where $|\mathbf{p}_1|$ is the magnitude of the three-momentum of particle 1 in the c.m. frame of the initial state, and the integration region is given by $\cos{\theta_1}\in[-1,1]$ and $\phi_1\in[0,2\pi)$.

For Fig.~\ref{fig_TwoDecEX_a}, using the corresponding rules, one has the following building blocks:
\begin{equation}
\begin{split}
    &V_1:~ \der\Phi_2(m;m_1,m_{(2)}),\hspace{.3cm}V_2:~ \der\Phi_2(m_{(2)};m_2,m_{(3)}),~\ldots,\hspace{.3cm}V_{n-1}:~ \der\Phi_2(m_{(n-1)};m_{n-1},m_n),\\
    &``(2)\text{''}:~(2\pi)^3\der m^2_{(2)},\hspace{.8cm}``(3)\text{''}:~(2\pi)^3\der m^2_{(3)},~\ldots,\hspace{0.8cm}
    ``(n-1)\text{''}:~(2\pi)^3\der m^2_{(n-1)}.
    \notag
\end{split}
\end{equation}
Multiplying them together and using Eq.~\eqref{eq_1to2PSFcm}, an expression for the $n$-body phase space element can be easily obtained,
\begin{equation}
    \der\Phi_n(m;m_1,\ldots,m_n)
    =\frac{|{\mathbf p}_1|\der\Omega_1}{2^{n}(2\pi)^{3 n} m} \prod_{i=2}^{n-1} |{\mathbf p}_{i}|\der\Omega_i\der m_{(i)}^{},
\end{equation}
where $(|\mathbf{p}_i|, \Omega_i)$ is the three-momentum of the final-state particle $i$ in the c.m. frame of the ($i,i+1,\ldots,n$) particle system. The integration region for the invariant mass $m_{(i)}$ is $\left[\sum^{n}_{k=i}m_k,m_{(i-1)}-m_{i-1}\right](i=2,3,\ldots,n-1)$ with $m_{(1)}=m$.

For the 3-body phase space, in addition to the chain tree diagram in Fig.~\ref{fig_TwoDecEX_a}, which reduces to the integration over a single invariant mass and angular variables, there is also a 1-loop diagram shown as Fig.~\ref{fig_TwoDecEX_b}. One has the following building blocks:
\be
\begin{split}
    &V_1:~ \der\Phi_2(m;m_3,m_{12}),\hspace{.5cm} V_2:~ \der\Phi_2(m;m_1,m_{23}),\hspace{.5cm}
     V_{3}:~ \der\Phi_2(m_{23};m_2,m_3),\\
    &``12\text{''}:~(2\pi)^3\der m^2_{12},\hspace{.5cm}``23\text{''}:~(2\pi)^3\der m^2_{23},\hspace{.5cm}
    \text{dashed single line ``3''}:~ \dfrac{(2\pi)^3}{\mathrm{d}^4p_3\delta(p_3^2-m_3^2)\theta(p^0_3)}.
    \notag
\end{split}
\ee
Multiplying them together leads to
\begin{align}
    \der\Phi_3(m;m_1,m_2,m_3)
    =\der\Phi_2(m;m_3,m_{12}) (2\pi)^3\der m_{12}^2 \der\Phi_{2}(m;m_1,m_{23}) \mathrm{d}m^2_{23}
    \hspace{.1cm}\delta\left[(p_{23}-p_3)^2-m_2^2\right].
\label{eq_1to3PSDloop}
\end{align}
Substituting Eq.~\eqref{eq_1to2PSFcm} into Eq.~\eqref{eq_1to3PSDloop}, one gets
\begin{align}
    \der\Phi_3(m;m_1,m_2,m_3)
    &= \frac{|\mathbf{p}_1||\mathbf{p}_3|}{(2\pi)^9 16m^2} \der\Omega_3\der m_{12}^2\der\Omega_1\der m_{23}^2
    \hspace{.1cm}\delta\left[m_{23}^2+m_3^2-m_2^2-2p_{23}^0p_3^0-2|\mathbf{p}_1||\mathbf{p}_3|\cos{\theta_3}\right] \notag\\
    &=\frac{1}{(2\pi)^9 32m^2}\hspace{.2cm}\der m_{12}^2\der m_{23}^2\der\Omega_1\der\phi_3
    \hspace{.1cm}\theta_{[-1,1]}\left(\frac{m_{23}^2+m_3^2-m_2^2-2p_{23}^0p_3^0}{2|\mathbf{p}_1||\mathbf{p}_3|}\right),
\label{eq_1to3PDFcm}
\end{align}
where all the integral variables are in the c.m. frame of the initial state. The effect of the boxcar function, which is defined as $\theta_{[-1,1]}(x)=1$ for $x\in[-1,1]$ and 0 otherwise, is to limit the invariant masses $m_{12}$ and $m_{23}$ to the physical region.

The 4-body phase space has more possibilities. In addition to the chain tree diagram, let us discuss two other diagrams shown as Fig.~\ref{fig_TwoDecEX_c} and \ref{fig_TwoDecEX_d}.
For Fig.~\ref{fig_TwoDecEX_c}, one has the following building blocks:
\be
\begin{split}
    &V_1:~ \der\Phi_2(m;m_{12},m_{34}),\hspace{.5cm} V_2:~ \der\Phi_2(m_{12};m_1,m_2),\hspace{.5cm}
    V_{3}:~ \der\Phi_2(m_{34};m_3,m_4),\\
    &``12\text{''}:~(2\pi)^3\der m^2_{12},\hspace{.5cm}``34\text{''}:~(2\pi)^3\der m^2_{34},
    \notag
\end{split}
\ee
and obtains using Eq.~\eqref{eq_1to2PSFcm},
\begin{align}
    \der\Phi_4(m;m_1,m_2,m_3,m_4)
    =\frac{|\mathbf{p}'_{12}||\mathbf{p}''_1||\mathbf{p}^*_3|}{(2\pi)^{12} 16m}\der m_{12}\der m_{34}\der\Omega_{12}'\der\Omega_1''\der\Omega_3^*,
    \label{eq_1to4Tree}
\end{align}
where $(|\mathbf{p}'_{12}|, \Omega_{12}')$ is the three-momentum of the final-state (1,2) particle system in the c.m. frame of the initial state, $(|\mathbf{p}''_{1}|, \Omega_{1}'')$ is the three-momentum of particle 1 in the c.m. frame of particles 1 and 2, and $(|\mathbf{p}^*_3|, \Omega_3^*)$ is the three-momentum of particle 3 in the c.m. frame of particles 3 and 4. The integration regions of $m_{12}$ and $m_{34}$ are $[m_1+m_2,m-m_3-m_4]$ and $[m_3+m_4,m-m_{12}]$, respectively.

For Fig.~\ref{fig_TwoDecEX_d}, the building blocks are
\be
\begin{split}
    &V_1:~ \der\Phi_2(m;m_{12},m_{34}),\hspace{.3cm} V_2:~ \der\Phi_2(m_{12};m_1,m_2),\hspace{.3cm}
    V_{3}:~ \der\Phi_2(m_{234};m_4,m_{23}), \\
    & V_{4}:~ \der\Phi_2(m_{23};m_2,m_3),\hspace{.3cm}``12\text{''}:~(2\pi)^3\der m^2_{12},\quad ``34\text{''}:~(2\pi)^3\der m^2_{34},\quad
    ``23\text{''}:~(2\pi)^3\der m^2_{23},\\
    &\text{dashed single line ``2''}:~ (2\pi)^{3}\left[\mathrm{d}^4p_2\delta(p_2^2-m_2^2)\theta(p^0_2)\right]^{-1}.
    \notag
\end{split}
\ee
Multiplying them together leads to
\begin{align}
 \der\Phi_4(m;m_1,m_2,m_3,m_4)
    =\,&\der\Phi_2(m;m_{12},m_{34}) (2\pi)^3\der m_{12}^2\der\Phi_{2}(m_{12};m_1,m_2) (2\pi)^3\der m_{34}^2 \notag\\
    & \times (2\pi)^3\der m_{23}^2 \der\Phi_{2}(m_{234};m_4,m_{23})
    (2\pi)^{-3}\delta\left[(p_{23}-p_2)^2-m_3^2\right].
\label{eq_1to4PSDloop}
\end{align}
Using Eq.~\eqref{eq_1to2PSFcm}, one can get
\begin{align}
    \der\Phi_4(m;m_1,m_2,m_3,m_4)
    =& \frac{|\mathbf{p}'_{12}||\mathbf{p}''_1|}{(2\pi)^{12} 2^7m m_{12}m_{234}|\mathbf{p}^*_2|}\der m_{12}^2\der m_{34}^2\der m_{23}^2\der\Omega_{12}'\der\Omega_1''\der\phi_4^*  \notag\\
    &\times\theta_{[-1,1]}\left(\frac{m_{23}^2+m_2^2-m_3^2-2p_{23}^{*0}p_2^{*0}}{2|\mathbf{p}_2^*||\mathbf{p}_4^*|}\right),
\end{align}
where $(|\mathbf{p}'_{12}|, \Omega_{12}')$ and  $(|\mathbf{p}''_{1}|, \Omega_{1}'')$ are as those in Eq.~\eqref{eq_1to4Tree}, and the quantities labelled by a ``$^*$'' are defined in the c.m. frame of (2,3,4) particle system in the final state. The invariant mass of particles 2, 3 and 4, $m_{234}$, is a function of $m_{12}$, $m_{34}$, $\Omega_{12}'$ and $\Omega_1''$.
The integration regions of $m_{12}$ and $m_{34}$ are $[m_1+m_2,m-m_3-m_4]$ and $[m_3+m_4,m-m_{12}]$, respectively, and the integration region of $m_{23}$ is limited by the boxcar function.

\section{Material 3: Graphic method in general spacetime dimensions}

The graphic method is also applicable in the case of more general $D$-dimensional spacetime, which is useful when considering dimensional regularization and quantum field theory in arbitrary dimensions.
The drawing rules and topological rules of the $D$-dimensional case are exactly the same as the $4$-dimensional case, and one only needs the following modifications (here one time dimension is considered):
\begin{itemize}%[leftmargin=*]
% \centering
    \item[(1)] \hspace{1.45cm}$(2\pi)^3\der m_{j_1j_2\cdots j_l}^2\to(2\pi)^{D-1}\der m_{j_1j_2\cdots j_l}^2$;
    \item[(2)] $\dfrac{(2\pi)^3}{\mathrm{d}^4p_j^{}\delta(p_j^2-m_j^2)\theta(p_j^0)}\to\dfrac{(2\pi)^{D-1}}{\mathrm{d}^D p_j^{}\delta(p_j^2-m_j^2)\theta(p_j^0)}$.
\end{itemize}
The $n$-body phase space element in $D$ dimensions is
\begin{equation}
\mathrm{d}\Phi_n(m;m_1,\ldots,m_n) = \delta^D(p-\sum_{i=1}^n p_i) \prod_{j=1}^n \frac{\mathrm{d}^{D-1}\mathbf{p}_j}{(2\pi)^{D-1} 2p^0_j}.
\label{eq_NBodyPSE_Dim}
\end{equation}
After integrating over the Dirac $\delta$-functions, there are $(nD-n-D)$ integral variables. The total number of independent two-body invariant masses is $C_n^2-1 = n(n-1)/2-1$, which gives the maximal value of $d$, the number of double lines $d$ in a diagram, for $n\leq D$. When $n\geq D-1$, the contributed number of Euler angles is given by the number of generators of the SO($D-1$) group, $(D-1)(D-2)/2$, and then the maximal value of $d$ is given by ${(D-1)(2n-D)}/{2}-1$.

The $2$-body phase space element with solid angle as the integral measure in $D$-dimension in any reference frame reads
\begin{align}
    \der\Phi_2(m;m_1,m_2) = &\sum_{|\mathbf{p}_1|}\frac{\der\Omega^{(D)}_1}{(2\pi)^{2D-2}} \frac{|\mathbf{p}_1|^{D-2}/4}{|(p^0|\mathbf{p}_1|-p_1^0|\mathbf{p}|\cos{\theta^{(D)}_{01}})|} \hspace{.1cm}\theta\left(p^0-\sqrt{|\mathbf{p}_1|^2+m_1^2}\right),
    \label{eq_2body_Dim}
\end{align}
where
\be
 \der\Omega^{(D)}_1=\der\phi_1\der\theta_{1,1}\cdots\der\theta_{1,D-3}\,\sin^{D-3}{\theta_{1,1}}\cdots\sin{\theta_{1,D-3}}
\ee
is the solid angle element of particle-1 in $D$-dimension space-time, with $\theta_{1,1},\ldots,\theta_{1,D-3}\in [0,\pi]$ the polar angles and $\phi_1\in[0,2\pi)$ the azimuth angle.
$\theta^{(D)}_{01}$ is the relative angle between the momentum of the initial state and that of particle-1 in the final state, which is given by
\begin{align}
    \cos\theta_{01}^{(D)} =&\, \cos\theta_{0,1}\cos\theta_{1,1} +
    \sin\theta_{0,1}\sin\theta_{1,1}\cos\theta_{0,2}\cos\theta_{1,2}
     + \ldots
    + \cos(\phi_1-\phi_0)\prod_{i=1}^{D-3}\sin_{0,i}\sin_{1,i} ,
\end{align}
with $\theta_{0,i}$ and $\phi_0$ the polar and azimuth angles of the initial state momentum, respectively.

\end{widetext}


\begin{thebibliography}{99}


\bibitem{Zyla:2020zbs}
P.~A.~Zyla \textit{et al.} [Particle Data Group],
``Review of Particle Physics,''
PTEP \textbf{2020}, 083C01 (2020).
% doi:10.1093/ptep/ptaa104

  %\cite{Jing:2019cbw}
 \bibitem{Forte:2002ni}
S.~Forte and G.~Ridolfi,
``Renormalization group approach to soft gluon resummation,''
Nucl. Phys. B \textbf{650}, 229 (2003)
% doi:10.1016/S0550-3213(02)01034-9
[arXiv:hep-ph/0209154].

 \bibitem{Jing:2019cbw}
H.-J.~Jing, S.~Sakai, F.-K.~Guo and B.-S.~Zou,
``Triangle singularities in ${J/\psi\rightarrow\eta\pi^0\phi}$ and ${\pi^0\pi^0\phi}$,''
Phys. Rev. D \textbf{100}, 114010 (2019)
% doi:10.1103/PhysRevD.100.114010
[arXiv:1907.12719 [hep-ph]].
%2 citations counted in INSPIRE as of 04 May 2020

\end{thebibliography}
\end{document}